\documentclass[journal,twocolumn]{IEEEtran}  
  
\usepackage{graphics}  
\usepackage{epstopdf}  
\usepackage[dvips]{graphicx,epsfig}  
\usepackage[usenames]{color}  
\usepackage{amsmath, amssymb, latexsym}  
\usepackage[makeroom]{cancel}  
\usepackage{xcolor} 
  
\begin{document}  
  
\title{Revisiting the Memristor Concept within Basic Circuit Theory}  
  
\author{Bernardo Tellini,~\IEEEmembership{Senior Member,~IEEE,}  
        ~Mauro~Bologna, Kristopher~J.~Chand\'ia,~\IEEEmembership{Member,~IEEE,} and Massimo~Macucci~\IEEEmembership{Member,~IEEE}  
\thanks{B. Tellini is with the Dipartimento di Ingegneria dell'Energia, dei Sistemi, del Territorio e delle Costruzioni, University of Pisa, Pisa I-56122, Italy (e-mail: bernardo.tellini@unipi.it)}  
\thanks{M. Bologna and K. Chand\'ia are with the Departamento de Ingenier{\'{\i}}a El\'ectrica - Electr\'onica, Universidad de Tarapac\'a, Arica, Chile (e-mail: mauroh69@gmail.com, kchandia@gmail.com).}  
%\thanks{K. Chand\'ia is with the  Escuela Universitaria de Ingeniería El\'ectrica - Electr\'onica, Universidad de Tarapac\'a, Arica, Chile (e-mail: kchandia@gmail.com).}  
\thanks{M. Macucci is with the Dipartimento di Ingegneria dell'Informazione, Universit\`a di Pisa, Pisa 56122, Italy (e-mail: macucci@mercurio.iet.unipi.it).}}  
  
\maketitle  
 
\begin{abstract}  
In this paper we revisit the memristor concept within
circuit theory. We start from the definition of the basic circuit elements,
then we introduce the original formulation of the memristor concept and
summarize some of the controversies on its nature.
We also point out the ambiguities
resulting from a non rigorous usage of the flux linkage concept.
After concluding that the memristor is not a fourth basic circuit element,
prompted by recent claims in the memristor literature, we
look into the application of the memristor concept to electrophysiology,
realizing that an approach suitable to explain the observed inductive behavior
of the giant squid axon had already been developed in the 1960s, with the
introduction of ``time-variant resistors.''
We also discuss a recent memristor implementation in which the magnetic flux
plays a direct role, concluding that it cannot strictly qualify
as a memristor, because its $v-i$ curve cannot exactly pinch at the origin.
Finally, we present numerical simulations of a few memristors and memristive
systems, focusing on the behavior in the $\varphi-q$
plane.
We show that, contrary to what happens for the most basic
memristor concept, for general memristive systems the $\varphi-q$ curve
is not single-valued or not even closed.
\end{abstract}  
\begin{IEEEkeywords}  
Circuit theory, Memristive Systems, Memristor, Numerical Simulations.  
\end{IEEEkeywords}  
  
\IEEEpeerreviewmaketitle  
  
\section{Introduction}  
  
\IEEEPARstart{T}{he} concept of memristor was first introduced by 
Chua~\cite{chua_71} in 1971 
as a two-terminal circuit element establishing a relationship  
between the charge 
(which is the integral of the current) and the integral of the voltage (which  
can also be defined as ``flux linkage''). Already in this pioneering paper,  
along with the detailed definition of the memristor and of its properties, a  
claim appeared that the memristor was to be considered as basic as the three  
classical circuit elements: the resistor, the capacitor, and the inductor. 
 
Here we focus on evaluating such a claim and on determining the essential  
nature of the memristor within circuit theory. 
%reaching the conclusion  
%that it is not a fourth basic circuit element, but rather an extension of  
%the classical concept of resistor, according to its defining equations and
%to the physical nature of its possible implementations. 
%In addition, we discuss a few examples of memristors (or, more in general,
%memristive systems) to show that effects that have been explained using 
%the memristor concept could be dealt with (and actually were 
%already comprehensively dealt with in the literature of the 60s) as 
%effectively with an extended resistor concept, as well as to provide further 
%evidence that any memristive behavior is associated with a dissipative 
%effect.
%
%We also include a few numerical simulations
%of memristive 
%devices, in order to clarify the actual properties (in particular in terms 
%of the $\varphi-q$ relationship) of such memristive devices. 
%  

The well-established basic elements in circuit theory are the inductor,   
the capacitor, and the  
resistor. Such elements introduce independent relationships between pairs  
of circuit quantities. In their most direct implementations, they can also 
be seen   as expressions of different physical laws.   
Citing the classical book by Desoer and Kuh~\cite{kuh},  
we report verbatim the definition of the inductor: {\it a   
two-terminal element will be called an  
inductor if at any time $t$ its flux $\varphi(t)$ and current $i(t)$ satisfy a  
relation defined by a curve in the $i,\varphi$ plane. The essential idea is 
that  
there is a relation between the instantaneous value of the flux $\varphi(t)$  
and the instantaneous value of the current}.   
It is important to point out that  
from the point of view of circuit theory and analysis the flux $\varphi(t)$ 
is the flux linkage, i.e.,  
the time integral of voltage, which is an actual  
circuit quantity.  
In the particular  
case of the classical idealized inductor obtained by winding an ideal thin 
conductor,
such an integral happens to coincide with the magnetic flux linked with the   
coil, which in turn corresponds to the actual magnetic flux piercing it only  
in the case of a coil with a single turn.  
For an inductor, the relationship between the time integral of the voltage and  
the current can be seen as an expression   
of Lenz's law, although this is not the only way to obtain such an   
``inductive'' behavior,  
which could well be implemented in a more elaborate fashion also in a world 
without magnetic field (for example with a capacitor and a gyrator).  

The capacitance establishes a relationship between the voltage $v(t)$ and the   
integral of the current (i.e., the charge $q(t)$) described by a curve in the  
$q,v$ plane. In the particular case of   
a classical capacitor consisting of two conductors separated by a   
dielectric, this is an expression of Gauss' law.   
Also in this case, we report   
verbatim from \cite{kuh}: {\it {The basic idea is that there is a relation  
between the instantaneous value of the charge $q(t)$ and the instantaneous  
value of the voltage $v(t)$.}}   

Finally, the resistance establishes a relationship between the voltage  
$v(t)$ and the current $i(t)$, through a curve on the $i,v$ plane. Linear   
resistors are implemented through Ohm's law, i.e., with conductors exhibiting   
a linear relationship between current and voltage.  
In \cite{kuh}, the authors write: {\it {The key idea of a resistor   
is that there   
is a relation between the instantaneous value of the voltage and the   
instantaneous value of the current}}.  
Also on the basis of the examples provided by Desoer and Kuh, this means that  
such a relation may be independent of the current and constant in time (linear  
time-independent resistor), dependent on the current but constant in time   
(non-linear time-independent resistor), independent of the current but time  
dependent (linear time-dependent resistor), or dependent on the current and  
on time (non-linear time-dependent resistor).    
  
Within a strict interpretation, this definition does not include many  
components  
of general use, which are commonly referred to as resistors because they are  
characterized by a dissipative behavior and the voltage at their terminals  
drops to zero whenever the current vanishes (which is equivalent to having
a ``pinched'' $v-i$ relationship, an expression commonly 
used in the memristor literature).  
  
A very simple example is represented by the incandescent light bulb, which is  
characterized by a resistance depending on temperature: since the temperature 
behavior exhibits a delay with respect to the current behavior (as  
a result of the thermal inertia of the filament) and depends on the ambient  
temperature, this would not fit any  
of the resistor definitions by Desoer and Kuh. We will discuss this issue   
in detail in Sec.~\ref{basic}.  
  
Summarizing, each basic element is associated with a single-valued curve  
relating two electrical quantities ($i$ and $v$ for the resistor, $i$ and  
$\varphi$ for the inductor, $v$ and $q$ for the capacitor). Such a curve  
is in general dependent on time (time-dependent elements), can be   
time-independent (time-independent elements), or can be just a line (linear  
elements, and, more in detail, a line with a slope depending on time   
for the linear time-dependent elements or just a single line for linear  
time-independent elements). 

To make our argument more rigorous, we provide a working definition
for the concept of basic circuit element: a basic circuit element, in its
simplest implementation, establishes a relationship between two circuit
quantities that cannot be reproduced by any single other basic circuit 
element. By simplest implementation we mean the linear and time-independent
case. For example, if two circuit elements are characterized by different 
physical dimensions, they certainly define independent relationships.
Conversely, in the case of a memristor with a constant value, 
we obtain the same relationship between flux linkage and 
charge as that implemented by a resistor.

In 1976 Chua and Kang~\cite{chua_76} extended the concept of memristor,  
indicating the memristor as a special case of a class of   
dynamic systems (memristive systems)   
for which the following relationships hold:  
  
\begin{eqnarray}  
 && {\dot{\vec x}} ={\vec f}({\vec x},u,t)  
\label{pie}
  \\   
  && y =g({\vec x},u,t)u   
   \label{torta}  
\end{eqnarray}  
being ${\vec x}$ the vector of the state variables of the system,  
$u$ a generic   
input quantity and $y$ a generic output quantity. In principle $y$ and $u$ 
can be  
arbitrary physical quantities.   
  
In the same 1976 paper, Chua and Kang defined a particular memristive system 
in which $y$   
is a voltage and $u$ is a current as a current-controlled memristive one-port.   
Analogously they defined a memristive system in which $y$ is a current and  
$u$ is a voltage as a voltage-controlled memristive one-port.  
  
In 2009 Di Ventra {\sl et al.}~\cite{diventra_2009} analyzed two particular   
types of memristive  
systems, one with $y=q$ and $u=v$ and the other with $y=\varphi$ and   
$u=i$, defining them memcapacitive systems and meminductive systems,   
respectively. A Lagrangian-Hamiltonian formalism for the description of   
such devices was then proposed by Cohen {\sl et al.}~\cite{diventra_prb} in   
2012.  
  
More recently, in 2015, Chua~\cite{everything} provided a detailed   
overview, reporting a hierarchy going from a class that is slightly less   
general than the memristive one-port down to the original 1971 memristor   
(reclassified as ``ideal memristor'').   
  
For the purposes of our discussion, we will not need this more detailed   
classification (which does not add new contributions to the basic memristor
theory, but it is just a more detailed taxonomy) and will refer to the 
definitions of the 1971 and 1976 papers,  
limiting ourselves to current-controlled and voltage-controlled   
memristive one-ports. 
It is our impression that the 1971 and the 1976 definitions of 
memristor and memristive systems, respectively, along with the extension to
the memcapacitor and meminductor, represent a very clear 
framework, and that the introduction of further distinctions and/or extensions
is not strictly necessary. For example, the ``extended memristive device'' 
introduced
by Valov {\sl et al.}~\cite{valov} is indeed just a circuit in which
a memristive system (according to the 1976 definition), a nonlinear 
resistor (according to the definition by Desoer and Kuh) and a voltage
source are present. 
In addition, we believe that it would be convenient to use the word 
``memristor'' only for the device defined in the 1971 paper by Chua,
avoiding its usage for the more general ``memristive systems'' (such 
as, for example, the light bulb).
Indeed, as we will discuss in Section III, general memristive systems are
not guaranteed to exhibit a single-valued $\varphi-q$ characteristic, which was one of the fundamental properties of the original 1971 memristor.  
  
Memristors have received significant attention by the scientific community,  
both for the analysis of circuit behavior and for possible industrial   
applications~\cite{corinto18},\cite{tse},\cite{tetzlaff},\cite{maldonado}.   
Such an interest into the memristor was mainly triggered by the 2008   
paper by   
Strukov $et$~$al.$~\cite{strukov}, in which experimental data were reported  
showing a behavior of a memristive nature for a nanoscale $\text{TiO}_2$   
device.  
  
In more detail, a titanium oxide film with a thickness of 5~nm was  
sandwiched between two metal (platinum) electrodes and was made up of two  
layers: an insulating $\text{TiO}_2$ layer and a conducting  
$\text{TiO}_{2-x}$  
layer (where conduction is due to oxygen vacancies acting as dopants).  
Such vacancies drift in the presence of an electric field, thereby  
moving the boundary between the conducting and the insulating layer and  
thus varying the overall resistance.  
  
In \cite{vongehr}, the authors criticize the interpretation of such a  
device as a memristor on the basis that a ``real memristor'' should involve  
magnetic flux. They also state the opinion that the original  
hypothesized memristor is still missing and probably impossible, while  
recognizing that a real memristor may in principle be discovered.  
  
The analysis of the memristor nature that we will present in the following   
leads instead to the conclusion that the $\text{TiO}_2$ device can well be  
considered a memristor (since, as it will become apparent from our  
discussion, the magnetic field is not essential, or even relevant, in the 
definition of the  memristor).  
  
In \cite{abraham} Abraham maintains that the device  
proposed in \cite{strukov} is not Chua's postulate of 1971, while also   
rejecting the memristor as a new basic circuit element.   
He sets out to demonstrate the non basic nature of the  
memristor with an involved argument based on a periodic   
table of basic elements and on   
two rules; one inferred from an analogy with the periodic table of   
chemical elements and the other one from the assumption that the definition   
of basic elements should be given outside transient conditions.   
Overall, Abraham points out some of the fundamental problems in the original  
memristor discussion (such as the ambiguity between the actual magnetic  
flux and the time integral of the voltage) and provides insightful hints  
into the nature of the memristor, without, however, making a 
complete case.  
  
Our analysis, once we have reached in a more direct way the same conclusion   
as Abraham's that the memristor is not a fourth basic circuit element,   
focuses on the intimate nature of a current or voltage-controlled   
memristive one-port. To this purpose, we provide an extended revisitation and   
analysis of significant past literature, and discuss the detailed behavior  
of a few relevant memristive one-ports, with a special focus on the 
associated   
curves in the $\varphi-q$ plane.   
Our final conclusion is that the current-controlled or voltage controlled  
memristive one-ports, and, specifically, the memristor (``ideal'' memristor  
defined in \cite{chua_71}) can be seen just as    
generalized resistors. Such a concept was indeed already proposed by 
Alexander Mauro~\cite{mauro} in 1961, in order to explain the appearance of 
a reactive component in the differential impedance of the giant squid axon 
and of the thermistor.  
  
The structure of the paper is as follows: 
Sec. II contains the main discussion on the basic nature of the memristor 
and of relevant examples from the literature. It is divided into 5 subsections:
in Subsection A we summarize the main defining memristor properties; in 
Subsection B we provide a clarification about the issue of the flux linkage
vs. magnetic flux; in Subsection C we present the reasons why we believe that
the memristor is not a basic circuit element; in Subsection D we focus on 
two enlightening examples in which the appearance of a reactive component
of the impedance is explained with the introduction of a generalized resistor
concept; and, finally, in Subsection E we perform an analysis of a recently
published memristor implementation, pointing out that its $v-i$ characteristic
is not rigorously pinched. 
In Sec. III, we present a few numerical examples useful to clarify the   
properties of memristors vs. those of more general memristive systems in   
the $\varphi-q$ and $v-i$ planes. In particular, we focus on the conditions  
needed to obtain a closed or even single-valued (as in the case of the   
original memristor definition) curve.  
Finally, we present our conclusions.  
  
\section{Basic Discussion}  
\label{basic}  
\subsection{Memristor properties}
Let us start from the analysis of the memristor concept that was presented   
by Chua in his original work~\cite{chua_71}: a passive element characterized   
by a relationship of the type $f(q,\varphi)=0$. In that paper, he also made  
the distinction between   
a charge-controlled memristor and a flux-controlled memristor for which    
$f(q,\varphi)$ can be expressed as a single-valued function of the   
charge $q$ or the flux linkage $\varphi$, respectively. Thus, for example,   
for   
the simplest form of charge-controlled memristor, we have:  
  
\begin{equation}  
 \varphi=f_M(q)   
 \label{bacco}  
\end{equation}  
and the voltage across the memristor is:  
  
\begin{equation}  
  v(t)=M(q(t))i(t)\, .  
  \label{memristance}  
\end{equation}  
The term $M(q)$ is defined memristance and can be expressed as:
\begin{equation}
M(q)=\frac{d f_M(q)}{d q}\, .
\label{one}
\end{equation} 

Thus it is actually a differential  
memristance, because it establishes a proportionality relationship   
between the differentials of the original quantities ($\varphi$ and $q$) it
is supposed to relate. This is not consistent with the definitions for 
resistors, capacitors and inductors which directly establish a proportionality 
relationship between $v$ and $i$, $q$ and $v$, and $\varphi$ and $i$, 
respectively. 

Thus, in order to be perfectly consistent with the definitions for
the basic circuit elements, the memristance should instead have been defined as
the ratio of the flux linkage (the time integral of the voltage) to the 
charge (the time integral of the current), i.e., it should coincide with
the $M'$ in the relationship
\begin{equation}
 \varphi(t)=\int_{-\infty}^t v(\tau) d\tau = M' \int_{-\infty}^t i(\tau)
d\tau= M' q(t)\, 
\end{equation}

As acknowledged by Chua himself~\cite{chua_71}, if $M$ has no dependence on 
$q$, the memristor becomes exactly equivalent 
to an ordinary linear and time-independent resistor.  
 
Chua also pointed out~\cite{chua_71} that the existence of a single-valued 
relationship between the flux linkage $\varphi$ (time integral of the
voltage) and the charge $q$ was
at the basis of the memristor definition, in analogy with the relationships
mentioned by Desoer and Kuh for the basic circuit elements. He even presented
a design for a $\varphi-q$ tracer for the experimental acquisition of such 
a characteristic.

Let us now focus on the more general case of current (voltage) controlled
memristive systems, defined by (\ref{pie}) and (\ref{torta}) when $u$ is a 
current and $y$ is a voltage ($u$ is a voltage and $y$ is a current).

The properties that they are expected to 
exhibit are:
\newline  
a) In \cite{kim} the authors report that the $v-i$ relationship shows a 
pinched hysteresis loop for a periodic 
input signal or equivalently, as stated by Chua in ~\cite{chua_2015}, all
memristive systems are characterized by the property that a zero output 
corresponds to a zero input, i.e., that for $i=0$ we always have $v=0$;
\newline 
b) In \cite{ascoli_2018}, Ascoli {\it et al.} state: {\it The memristor ... 
is a two-terminal nonlinear dynamic circuit element obeying a Ohm law 
dependent upon the time evolution of its memory state.} 
\newline
c) For the particular case of $M$ being only a function of $q$, the area 
of the hysteresis loop decreases with frequency and the hysteresis loop
collapses to a single-valued function in the limit of $f\to\infty$, being $f$
the operating frequency. 
  
Any of these properties is consistent with 
the memristive system being simply a resistor whose value depends on the 
internal state vector $\vec x$, on the instantaneous value of the current 
(for the current-controlled case) and, possibly, on time.   

However this does not just correspond to a time-dependent resistor, because,
in the commonly accepted interpretation within circuit theory, a quantity
is time-dependent if its value has a direct dependence on time, i.e., its
partial derivative with respect to time is non zero and there is no other
indirect dependence on time.

In the case of the memristor, instead, the memristance depends on variables
which are in turn functions of time, therefore it has a zero partial derivative
with respect to time (in the case of a ``time-independent'' memristor), but 
a non-zero total derivative with respect to time.

In 1961 Mauro~\cite{mauro} introduced the concept of a ``time-variant'' 
resistor, an element whose resistance ``varies intrinsically with time by 
virtue of the fact that the physical state of the element changes with time,'' 
while a time-dependent resistor is an element whose resistance varies
``as an independent function of time'' (in accordance with the
definition by Desoer and Kuh).
%Property c) implies that, for very high frequency, the hysteresis loop becomes
%a straight line and therefore we have again the same behavior as that of 
%a basic resistor.
%
In Subsections C and D we provide a detailed discussion of this concept.

\subsection{Magnetic flux and flux linkage}
As already mentioned, $\varphi$ is just the time integral of the 
voltage  
(i.e., the flux linkage), which is a relevant quantity in circuit analysis
(while the magnetic flux is not an actual circuit quantity, except
in the case of ideal inductors, when it coincides with or is proportional to 
the flux linkage). 
This is fully consistent   
with the definition given by Chua in the first page of his original   
memristor paper of 1971~\cite{chua_71}. However, in Sec.~4 of the very same   
paper, Chua, trying to provide a more general interpretation of the memristor,  
in terms of a quasi-static solution of Maxwell's equations, states   
that the surface integral of the magnetic flux density  
is proportional to the flux linkage. The problem is that such a flux linkage
(originally defined relative to an inductor and equal to the magnetic
flux linked by each loop of the inductor multiplied by the number of 
loops) is not, except for the case of an ideal inductor,  
the flux linkage used in the first part of the paper for the 
definition of the memristor, i.e., 
the time integral of voltage.  
  
This discrepancy is at the origin of the ambiguity (integral  
of the voltage vs. actual magnetic flux) in the meaning  
of $\varphi$ that may have contributed to some involved arguments on  
the nature of the memristor which can be found in the literature and to
the search for a memristor relating the actual magnetic flux with the
charge, which turns out to be a somewhat ill-posed problem: in order to 
make the integral of voltage proportional to the actual magnetic flux
an inductance must be present, which will then lead to a $\varphi - q$ 
curve that cannot be exactly pinched in the origin.

In some papers the meaning of flux linkage used in the original memristor
definition appears to have been completely forgotten, and the memristor
is presented as a circuit element relating charge and 
``magnetic flux''~\cite{tour,microtub}.

More in general, circuit theory is indeed a special case of Maxwell's   
equations only as long as we consider classical physics and we limit   
ourselves to electromagnetic phenomena. If other effects (e.g. chemical,  
thermal, etc.) are included or if quantum mechanical effects are   
taken into account, a multiphysics approach is needed. In the introduction 
we have already  
mentioned the case of the thermistor (which requires the inclusion of the   
heat transport equations) and of the axon (which requires the inclusion   
of ion transport equations). Another example can be that of quantum   
capacitance~\cite{luryi,quantum_dot,buett}, which arises as a consequence of   
the quantum confinement energy and the relationship between the chemical 
potential and the charge.

If we consider the proper definition of $\varphi$, 
the $\text{TiO}_2$ based   
device presented in \cite{strukov} can indeed be defined a memristor, as
we had anticipated, because  
it is a resistor whose value is a function of the charge that has flowed   
through it, and thus of the time evolution of the current, which makes it a   
specific type of time-variant resistor, exhibiting the memristor  
properties defined in \cite{chua_71}.  

\subsection{Matrix representation of circuit elements} 
\begin{figure} 
    \centering 
    \includegraphics[scale=0.35]{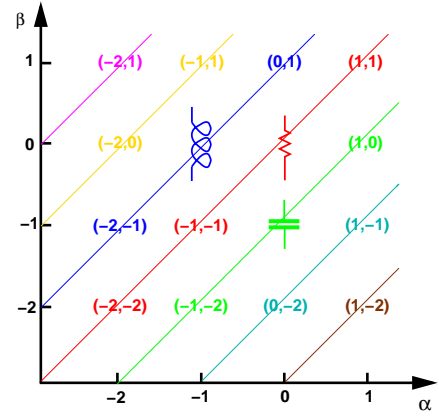}
    \caption{Mapping of circuit elements based on the orders of the 
derivatives of the voltage and of the current between which they
establish a relationship; there is a single basic element per
diagonal (adapted from Fig.~3 of \cite{chuatut1}).}
    \label{diag_fig} 
\end{figure} 

In \cite{chuatut1} Chua introduces a matrix representation 
of the circuit elements, which we summarize in Fig.~\ref{diag_fig}. In  
more detail, each element of the 
matrix can be identified by the pair $(v^\alpha, i^\beta)$, where  
$\alpha$ and $\beta$ indicate the column and row of the element, which  
correspond to the order of the time derivative of the voltage and of the 
current, 
respectively. An element located in the column $\alpha$ and in the row $\beta$ 
relates the time derivative of order $\alpha$ of the voltage to the time  
derivative 
of order $\beta$ of the current through a single-valued function. 
Thus, moving along the main diagonal, it is  
possible to identify the resistor and the memristor, which is in the position
($-1,-1$), while moving along the  
diagonal right below the main one we can identify the capacitor and  
the memcapacitor ($-1,-2$), and along the diagonal above the main one we 
find the  inductor and the meminductor ($-2,-1$)~\cite{diventra_2009}. 

If one were to assume that each element of the matrix corresponds to an 
independent basic circuit element, also the memcapacitor and the 
meminductor would be as basic as the resistor, the capacitor, 
the inductor, and the memristor. 

However, it is apparent that it cannot be so, because, first of all,
in the linear case all the elements of the same diagonal do coincide, 
which is by itself evidence that independent basic circuit elements
cannot exist along the same diagonal. Furthermore, considering the most
direct implementation, each diagonal is associated with a specific physical
law: Gauss's law for the diagonal below the main one, Ohm's law for the main
diagonal, and Lenz's law for the diagonal above the main one. Indeed also the
memristor, as the resistor, is in the end a dissipative element, and is 
therefore an expression of Ohm's law. The dissipative nature of the memristor
was realized also by Di Ventra and Pershin~\cite{divenpersh}, who point
out that, as a result of such dissipative behavior, the memristor violates
the time-reversal invariance.
This intrinsically resistive nature 
of the memristor has sometimes been overlooked because of the mistaken
association with the magnetic flux instead of with the time integral of the 
voltage.

The previously mentioned interpretation of $\varphi$ as the actual magnetic 
flux appears to be also at the origin of the choice by Wang
of the memristor,
the memcapacitor and the meminductor as the elements for his
``basic element triangle''~\cite{triangle} that he proposed after recognizing 
the inconsistency of assuming the memristor as a fourth basic circuit element. 
Such a formulation appears to be reasonable and free from the contradictions
of the four basic element assumption. It is also consistent with the position 
of the memristor, memcapacitor and meminductor in three different diagonals
of Fig.~\ref{diag_fig}.
Wang reaches this conclusion mainly because 
he considers the flux $\varphi$ more fundamental than the voltage, a 
somewhat understandable argument if $\varphi$ were the actual magnetic flux,
which however is not, as we have already pointed out.

In addition, as previously discussed, the memristance is
defined exactly as a resistance with more general properties, instead of the 
ratio of the flux linkage to the charge (which would still coincide with 
a resistor in the linear, time-independent case). The same is true for the 
memcapacitor, defined by $q=C({\vec x}, v, t) v$~\cite{divenpersh} 
(instead
of $\int q(t)dt=C' \int v(t) dt =C' \varphi$) and for the meminductor, 
defined by $\varphi =L({\vec x}, i, t) i$~\cite{divenpersh} (instead 
of $\int\varphi(t) dt 
= L' \int i(t) dt= L' q$).
%On the other hand, Wang's ``basic element triangle'' would have the advantage
%of being more general than the ``classical triangle,'' 
%since the memristor, the memcapacitor and the meminductor can be interpreted
%as generalizations of the resistor, the capacitor, and the inductor.

Thus it is far simpler to just generalize the definitions
of the circuit elements provided by Desoer and Kuh, by including time-variant 
elements, as proposed by Mauro for the resistor, and to keep the resistor,
the capacitor and the inductor as the basic circuit elements (which in this
way would also include all the properties of the memristor, memcapacitor and 
meminductor). 

%Furthermore the resistor 
%relates the two quantities, i.e., voltage and current, which are most 
%commonly used in circuit analysis, as, for example, in Kirchhoff's laws.  

%Such an association with the time integral of the voltage, as in
%the original 1971~\cite{chua_71} definition, 
%should always be used not to lose contact with the actual nature
%of the memristor. 

Possibly, novel basic elements could be associated to more external 
diagonals, even though their practical realization is apparently far from 
being conceived, 
and a discussion on this is outside the purpose of the present work.  
 
Overall our conclusion is that a fourth basic circuit element does not exist, 
and that the memristor can be seen simply as a generalization of the resistor, 
a time-variant resistor, according to the definition given by Mauro in 1961,
which we will discuss in detail in the next subsection. 
A further generalization of the time-variant resistor concept by Mauro,
introducing also a direct dependence on time, covers the most general
definition 
of a memristive current controlled (or voltage controlled) one-port
provided by Chua and Kang~\cite{chua_76}, in which the value of the 
memristive one-port is a generic function   
of a set of state variables, the current (or the voltage in the case of 
a voltage controlled memristive one-port), and, possibly,  
time. 

As long as we assume a reasonable boundedness condition for the memristance 
(as suggested by Chua in~\cite{pinched} and in~\cite{comcom} to prevent the
problems pointed out by Pershin and Di Ventra in~\cite{comment}), the fact
that the curve in the $i-v$ plane is pinched is simply equivalent to 
stating that there is no energy storage in the device and, thus, that the 
device is purely dissipative, i.e., a resistor.
 
%For the sake of completeness, we also point out that the ``memristor''  
%element on the main diagonal of Fig.~\ref{diag_fig} corresponds to the  
%original 1971 memristor~\cite{chua_71}, which defines a single-valued 
%curve in the  
%$\varphi-q$ plane, and not to generalized memristive devices, which  
%are not typically described by a single-valued curve in such a plane
%(as we will show in the next section). 
  
\subsection{Time-variant resistors}  
It is now appropriate to go back to the example we mentioned in the   
introduction: the filament of the incandescent light bulb, which is one of   
the simplest cases that do not fit into the textbook definition of   
resistor that we quoted at the beginning.  

Already in 1961, Alexander Mauro~\cite{mauro}, realized the difference 
between   
an incandescent light bulb and a time-dependent resistor, and defined it as a   
``time-variant'' resistor, as mentioned at the end of Subsection A. 

%According to Mauro, a time-variant resistor is an   
%element whose resistance ``varies intrinsically with time by virtue of the   
%fact   
%that the physical state of the element changes with time'', while a   
%time-dependent resistor is an element whose resistance varies ``as an 
%independent function of time'' (in accordance with the definition by Desoer 
%and Kuh).  

In particular, Mauro provided analytical evidence that a circuit containing   
a thermopositive element (i.e., an element whose resistance increases with  
temperature) such as an incandescent light bulb exhibits a differential 
impedance  of capacitive nature (as a consequence of the delay in the 
temperature   
response to current variations), while with a thermonegative element a   
differential impedance of inductive nature would appear. In the paper by
Mauro this was   
introduced to provide a physical justification of the so-called ``anomalous  
impedance,'' which had been observed in giant squid axons~\cite{cole}.  
Indeed, early measurements of the AC impedance of the giant squid axon had  
revealed a significant inductive component, which could not be explained   
as a result of magnetic energy storage, due to the absence of a large   
enough loop. This is the reason why Cole~\cite{cole} mentioned that this   
property had been found shocking, but he accepted it, and correctly   
attributed it to properties of the membrane (from the observation that it   
could not be associated with a flow of current parallel to the axon, due to   
the fact that an asymptotic inductance value was reached when increasing the   
distance between the electrodes along the axon).   
Mauro, with the case study of the temperature-dependent resistor, was able to   
present a clear example of how a time-variant resistor (a resistor whose   
value depends on state variables, which in turn depend on time) may exhibit   
a differential inductive behavior. Then in 1965 Cohen and Cooley~\cite{cohen}   
performed a numerical solution of the Nernst-Plank equations for ionic   
transport through a thin permeable membrane, finding, for this specific   
physical system, an inductive behavior, as shown in their Figs.~5   
and 6, analogous to that derived by Mauro for a generic thermonegative element.  
%Such an inductive response is due to the delay associated with ionic   
%diffusion.  
%  
Thus, by 1965 the presence of an inductive component in the impedance of a   
membrane, and therefore of an axon, had been understood and explained without
requiring the presence of any magnetic field.  
  
It is therefore surprising that in a recent paper~\cite{axon_chua}, the   
presence of the axon inductance is reported as a problem that had remained  
unsolved for 70 years and that for the same amount of time had been associated  
with an ``enormous magnetic field''.   
  
In general a magnetic field is not needed to produce   
an inductance-like behavior, because it can simply arise as a result of   
inertia: inertia of carrier motion for the kinetic inductance~\cite{meservey},  
thermal inertia for thermonegative elements, delay associated with ionic   
diffusion for the axon inductance, etc. It is however important to point out  
that not all of these three examples are of a memristive nature: while in the   
second and the last example   
there is no energy storage, in the first example energy is stored in   
the form of kinetic energy and can be returned to the circuit. Therefore  
in the case of kinetic inductance we are dealing with an actual inductance  
which is not associated with a pinched $i-v$ characteristic, while in the   
other two cases there is no energy restitution and we have a memristive   
behavior.  
  
The authors of \cite{axon_chua} redefine the time-variant resistors  
of the Hodgkin-Huxley~\cite{hodgkin} paper as time-invariant memristors 
(time-invariant memristive systems, on the basis of the definitions we  
decided to use in the introduction, since, as we will show in Sec. III, 
they are not characterized by a single-valued curve in the $\varphi-q$ 
plane) and then   
proceed to the calculation of the differential impedance of the overall   
circuit, finding the expected inductive component.   
  
However, the same  
calculation could have been performed without renaming them, and following, 
for  example, the approach used by Mauro to treat the temperature-dependent  
resistor. Furthermore,   
Mauro provided a clear physical explanation of the appearance of an inductance  
behavior, while in \cite{axon_chua} the inductance is the result of a   
long and involved analytical calculation, without direct physical insight.  
  
Thus, both definitions, time-variant resistor and memristive system, are  
acceptable and, if properly handled, lead to the same results.   

%Apart from this naming issue, the time-variant resistor or the memristive  
%system are representative of a dissipative process, and thus they do not  
%establish   
%a relationship independent of the basic one established by the 
%traditional resistor (as defined by Desoer and Kuh).  
%This is clearly demonstrated by the fact that, as mentioned in the  
%introduction, in the simplest case, of a time-independent and linear  
%memristor (whose value thus does not depend on time, on the state variables,  
%or on the current),   
%we have the exact equivalent of a time-independent and linear resistor.  
  
\subsection{Magnetic core memory device - $\Phi$ memristor}  
In relationship to the search for a memristor connecting directly   
the charge with the magnetic flux (instead of the time integral of 
the voltage), Wang {\sl et al.}~\cite{wang} have recently  
published a paper in which they present a device that they define   
a ``real memristor''. Such a device is based on a conducting wire going  
through a magnetic core: if a   
rectangular current pulse is applied to the wire, the resulting magnetic 
field rotates the core magnetization, which in turn leads to a variation 
in time of the magnetic flux, and thus to an induced voltage. This is
exactly the readout process of the old magnetic core memories~\cite{renwick}, 
with the only difference that, while in core memories two separate loops
were used (one to inject the current and one for the readout), 
Wang {\sl et al.} use a single loop. 
They find a relationship  
between the magnetic flux and the charge, and, in particular, a pinched  
current-voltage relationship (as typical for memristors). 

However they do neglect the voltage induced as a result of the geometrical 
self-inductance of the loop, i.e., the inductive effect associated with the 
magnetic flux variation directly produced by the input current. 
In the case of square current 
pulses, the authors state explicitly that they have filtered out 
(by reducing the oscilloscope bandwidth) the ``transient spike,'' due to 
such inductance, which they define as ``parasitic'' (the spike is indeed 
visible in the experimental data of their Fig.~10). If instead we consider 
a sinusoidal excitation, as in (12) of~\cite{wang}, there is another 
term to be included, again due to the geometrical loop inductance, which is 
in quadrature with the one of~\cite{wang}. Such inductance is indeed 
not ``parasitic,'' because it is the source of the very same connection 
between magnetic flux variation and induced voltage that leads to the 
in-phase term. The in-phase term is the result of the flux variation due 
to the magnetization lining up with the driving magnetic field created by the 
injected current, while the quadrature term is the result of the
flux variation due to the change of such a driving magnetic field. 
The alignment of the magnetization is the consequence of the dissipative 
effect connected with the Gilbert term~\cite{gilbert}
thus leading to a time-variant resistive component (the memristor-like
part).
Depending on 
the waveform and on the magnetic core characteristics, the in-phase term 
may be much larger than the quadrature term, but the ratio of their amplitudes 
must in any case be finite: if the inductive component should vanish, the 
same would happen for the dissipative ``memristive'' component.
Thus the $\varphi$-memristor is not rigorously a memristor, because its
characteristic $v-i$ curve cannot be exactly pinched. 

Furthermore, the authors of~\cite{wang} in their Fig.~9 report
curves for the voltage and the current as a function of time that are
not consistent with the plot in the $i-v$ plane and with the experimental
results at the top of Fig.~9. Indeed, in the plot reporting the time behavior
of the current and the voltage, the current changes sign exactly when the 
voltage pulse ends. In such a case, and in any other case in which the 
current is reversed after the voltage pulse ends (as in the experimental 
results of Fig. 10), the curve on the $i-v$ plane would not look like the
one reported, but would consist of a collection of segments without
a hysteresis cycle: there would be a horizontal segment corresponding to
the current step, then a vertical segment corresponding to the increase
and decrease of the voltage at constant current, and then again a horizontal
segment reaching the negative current value, followed by a behavior symmetric
with respect to that during the positive current pulse.

Curves with hysteresis such as the one reported by the authors of~\cite{wang} 
for the $i-v$ plane and in their experimental results
at top of their Fig.~9 are possible only if the current is reversed before
the voltage pulse ends (which is also consistent with the frequencies they
report for the experimental result and the duration of the voltage pulses
in Fig. 10).

%Even assuming the particular condition in which the current pulse ends
%at the same time as the voltage pulse, the latter should be perfectly 
%symmetric around its peak for the $\varphi-q$ curve to be single-valued.
%From the experimental data of Fig. 10 in ~\cite{wang} as well as from the
%old literature on magnetic core memories~\cite{renwick}, it is apparent that
%this is not the case.

In order to further clarify the properties of different types of 
memristive one-ports, 
in the next Section we present a few numerical examples for  
relevant cases: the thermistor, the axon, and a memristor (according to 
a strict interpretation of the 1971 definition by Chua) made up of  
resistors and charge controlled switches.
For all the examples, we report the associated $\varphi-q$ curve and
discuss the conditions under which such a curve is single-valued. 
\section{Memristive Circuit Models}  
\label{geppo}  
A memristive functionality can be implemented in many different ways, 
exploiting a variety of different physical systems, even relatively exotic 
ones~\cite{caravelli}, since all that is needed is a dissipative effect and
a mechanism that can vary the resistance under the control of a state variable.

Here we focus on the analysis of a few simple but meaningful examples.

We start with a numerical analysis of the thermistor, which is the first 
memristive device that was studied by Mauro in 1961 to provide an example  
of how a differential impedance with a reactive nature can appear in the 
absence of actual physical capacitors or inductors and of energy storage.  
We consider the thermistor model of \cite{chua_76}, which is based on  
the following equations, including 
the dependence of the thermistor resistance on temperature (which is  
the state variable), as well as the time evolution of temperature as a  
function 
of the heat capacitance, the thermal resistance to the outside environment, 
and the dissipated electrical power.  
 
\begin{eqnarray} 
&&v=R_0(T_0)\exp{\left[\beta\left(\frac{1}{T}-\frac{1}{T_{0}}\right)\right]}i, 
\\ 
&&p(t)=v(t)i(t)=\delta(T-T_0)+C\frac{dT}{dt},\\ 
&&\frac{dT}{dt}\!=\!-\frac{\delta}{C}(T-T_0)+\frac{R_0(T_0)}{C}\exp\!\!\left[\!\left(\frac{\beta}{T}\!-\!\frac{\beta}{T_{0}}\right)\!\right]\!i^2\!, 
\end{eqnarray} 
where $\delta=0.1$~mW/$^{\circ}$C, $R_0=8$~k$\Omega$, $T_0=298$~K,  
$\beta=3460$~K, $C=1$~mJ/K.

\begin{figure}
\centering
\includegraphics[scale=0.3]{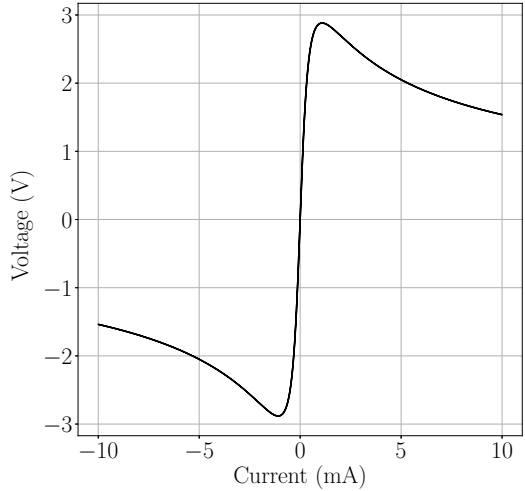}
\caption{Thermistor $v-i$ curve for $f=1$~$\mu$Hz,
$\delta=0.1$~mW/$^{\circ}$C, $R_0=8$~k$\Omega$, $T_0=298$~K, $\beta=3460$~K,
$C=1$~mJ/K.}
\label{vi_slow}
\end{figure}

We first report (Fig.~\ref{vi_slow}) the $v-i$ curve obtained by cycling the  
current in the low-frequency limit (1~$\mu$Hz), which corresponds to the 
result in Fig.~2 of \cite{chua_76}: in this case the hysteresis 
disappears and a single-valued relationship between voltage and current is 
established, since dynamical effects due to the delayed response of the  
state variable (the temperature) to the time behavior of the dissipated  
electrical power are negligible at such a low frequency, which is 
much smaller than 
the reciprocal of the relevant time constant of the system. 

\begin{figure}
\centering
\includegraphics[scale=0.25]{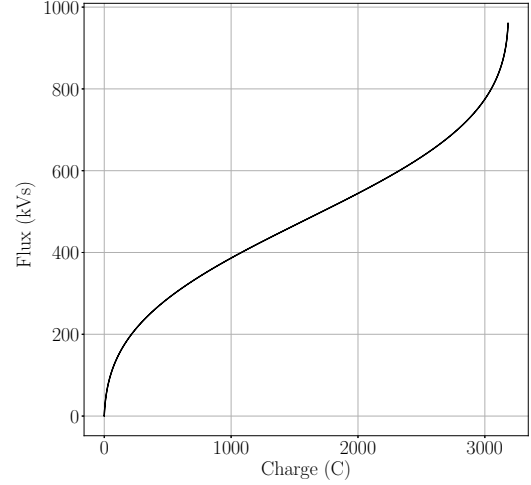}
\caption{Thermistor $\varphi-q$ curve for $f=1$~$\mu$Hz,
$\delta=0.1$~mW$/^{\circ}$C, $R_0=8$~k$\Omega$, $T_0=298$~K, $\beta=3460$~K,
$C=1$~mJ/K.}
\label{pq_slow}
\end{figure}

In Fig.~\ref{pq_slow}, we report the associated curve in the $\varphi-q$ 
plane, which, in this particular case without hysteresis, is single-valued. 
Therefore at vanishingly small frequency the thermistor does have a property  
(an unambiguous definition through a curve in the $\varphi-q$ plane) that was  
reported as a fundamental characteristic of a memristor in \cite{chua_71}.
Such an unambiguous relationship between the two defining quantities (e.g., 
voltage and current in the case of the resistor) is also at the basis of  
the basic circuit element definitions by Desoer and Kuh~\cite{kuh}. 

\begin{figure}
\centering
\includegraphics[scale=0.3]{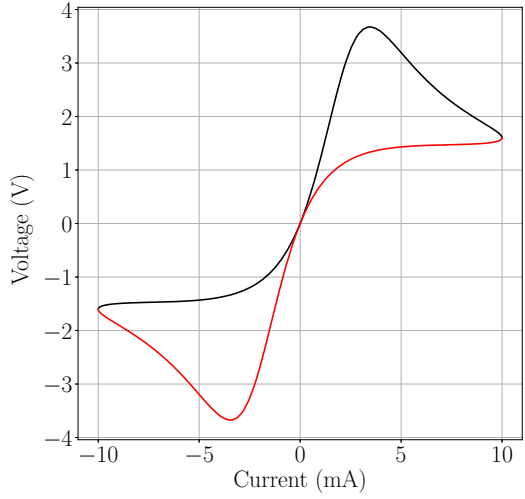}
\caption{Thermistor $v-i$ curve for $f=10$~mHz, $\delta=0.1$~mW$/^{\circ}$C,
$R_0=8$~k$\Omega$, $T_0=298$K, $\beta=3460$~K, $C=1$~mJ/K.}
\label{vi_medium}
\end{figure}

We then consider (Fig.~\ref{vi_medium}) the $v-i$ curve of the same 
device obtained cycling the current at a higher frequency (0.01~Hz):  
in this case a hysteresis 
appears, as a result of the delay of the thermal response. The 
red portion of the curve corresponds to the descending half-period of the 
current, while the black portion is relative to the ascending  
half-period. 
It is to be noted that neither the ascending nor the descending 
branches are symmetric around the origin, which implies that a different 
amount of flux linkage is transferred in the first and in the third quadrant. 

\begin{figure}
\centering
\includegraphics[scale=0.3]{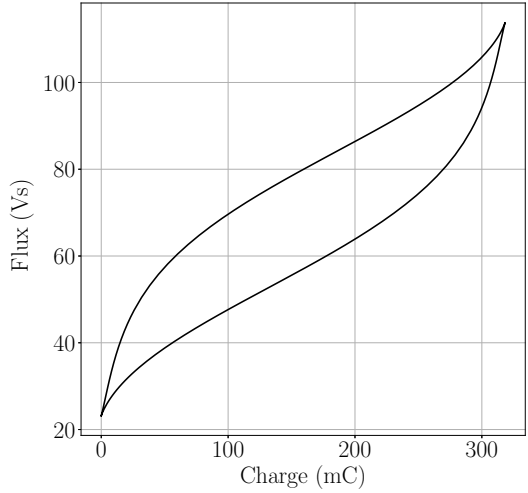}
\caption{Thermistor $\varphi-q$ curve for  $f=10$~mHz,
$\delta=0.1$~mW$/^{\circ}$C, $R_0=8$~k$\Omega$, $T_0=298$~K,
$\beta=3460$~K, $C=1$~mJ/K.}
\label{pq_medium}
\end{figure}

A complementary asymmetry is present in the other branch, which leads to  
a hysteresis in the $\varphi-q$ plane. The curve in the $\varphi-q$ plane
is not 
single-valued, neither with respect to $\varphi$ nor with respect to $q$,  
although it is closed (Fig.~\ref{pq_medium}). As a consequence, at such a 
frequency this thermistor model does not exhibit the properties of the 
memristor defined  
of 1971, although it still belongs to the more general category of  
time-variant resistors [and fits in the later definition (1976) of memristive
systems]. 

\begin{figure}
\centering
\includegraphics[scale=0.3]{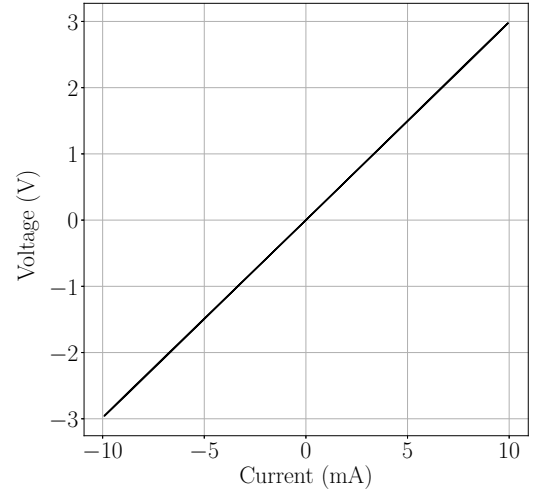}
\caption{Thermistor $v-i$ curve for $f=10$~Hz,
$\delta=0.1$~mW$/^{\circ}$C, $R_0=8$~k$\Omega$, $T_0=298$~K,
$\beta=3460$~K, $C=1$~mJ/K.}
\label{vi_fast}
\end{figure}

\begin{figure}
\centering
\includegraphics[scale=0.3]{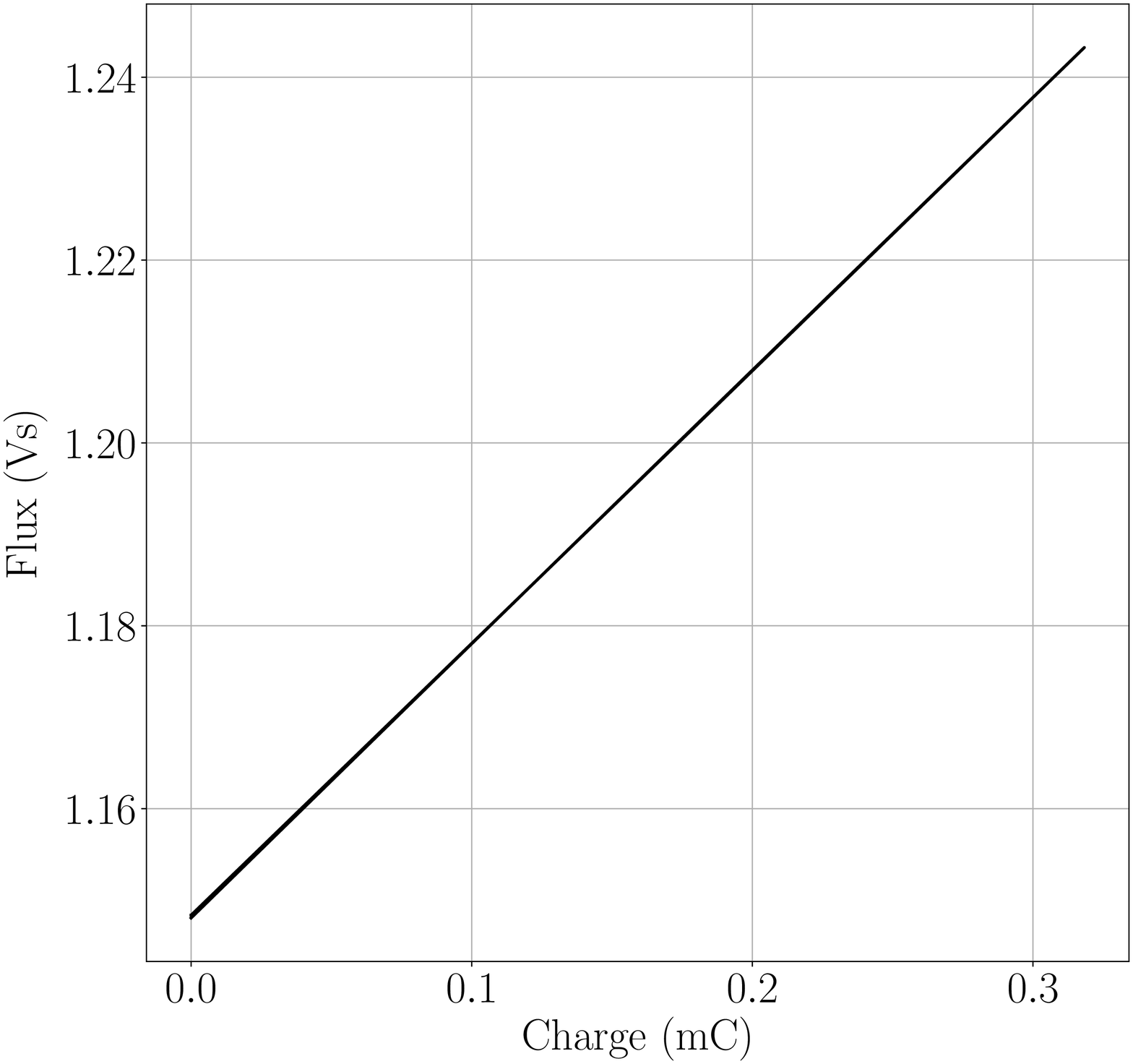}
\caption{Thermistor $\varphi-q$ curve for $f=10$~Hz,
$\delta=0.1$~mW$/^{\circ}$C, $R_0=8$~k$\Omega$, $T_0=298$~K,
$\beta=3460$~K, $C=1$~mJ/K.}
\label{pq_fast}
\end{figure}

If we cycle the current at a relatively high frequency (compared to the  
reciprocal of the characteristic time constant of the device), the hysteresis 
disappears and we obtain a substantially linear $v-i$ curve, with a slope  
determined by the value of the reached equilibrium resistance, as shown  
in Fig.~\ref{vi_fast}. Accordingly, also in the $\varphi-q$ plane we 
obtain a linear, single-valued curve (Fig.~\ref{pq_fast}).  

We now briefly study the circuit model for the potassium channel of the 
axon formulated by  
Hodgkin and Huxley~\cite{hodgkin}, in the form from \cite{axon_chua}.  
We have 
\begin{eqnarray}\label{om} 
&&i_K=G_K(x_1)v_K 
\\\label{cond} 
&&G_K(x_1)=g_K x_1^4 
\\\nonumber 
&&\frac{dx_1}{dt}=\left\{\frac{\gamma(v_K+E_K)} 
{\exp\left(\frac{v_K+E_K}{\hat V}-1\right)}\right\}(1-x_1) 
\\ \label{ve} 
&&-\frac{1}{\tau}\left[\exp\left(\frac{v_K+E_K}{8 {\hat V}}\right)\right]x_1 
\end{eqnarray} 
We performed numerical simulations with the parameters  
$g_K=36$~S, $E_K=10$~V,  
$\gamma=0.01$~V$^{-1}$s$^{-1}$, ${\hat V}=10$~V, $\tau=8$~s. 
 
\begin{figure}
\centering
\includegraphics[scale=0.3]{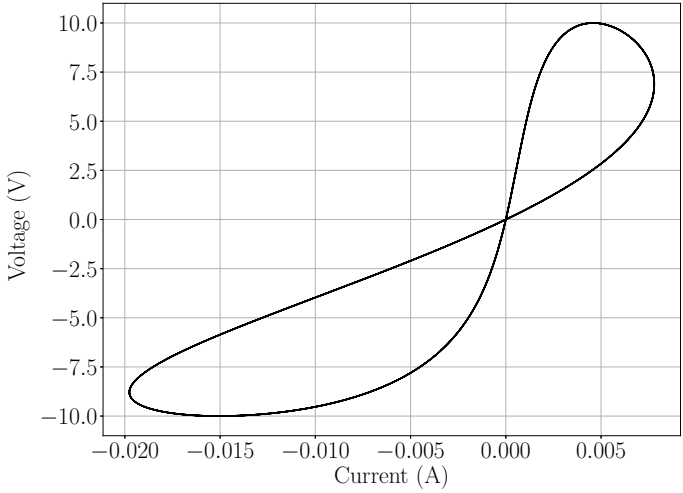}
\caption{Axon model $v-i$ curve for the following parameters:
$f=50$~mHz, $g_K=36$~S, $E_K=10$~V,
$\gamma=0.01$~V$^{-1}$s$^{-1}$, ${\hat V}=10$~V, $\tau=8$~s.}
\label{vi_axon}
\end{figure}

In Fig.~\ref{vi_axon} we report the $v - i$ curve obtained cycling at a  
frequency of 50~mHz, which is comparable with the reciprocal of the time 
constant $\tau$. We observe that it pinches in the origin, according 
to (\ref{om}), and that it is not symmetric around the origin, due to  
the particular nature of the defining equations. 

\begin{figure} 
\centering 
\includegraphics[scale=0.3]{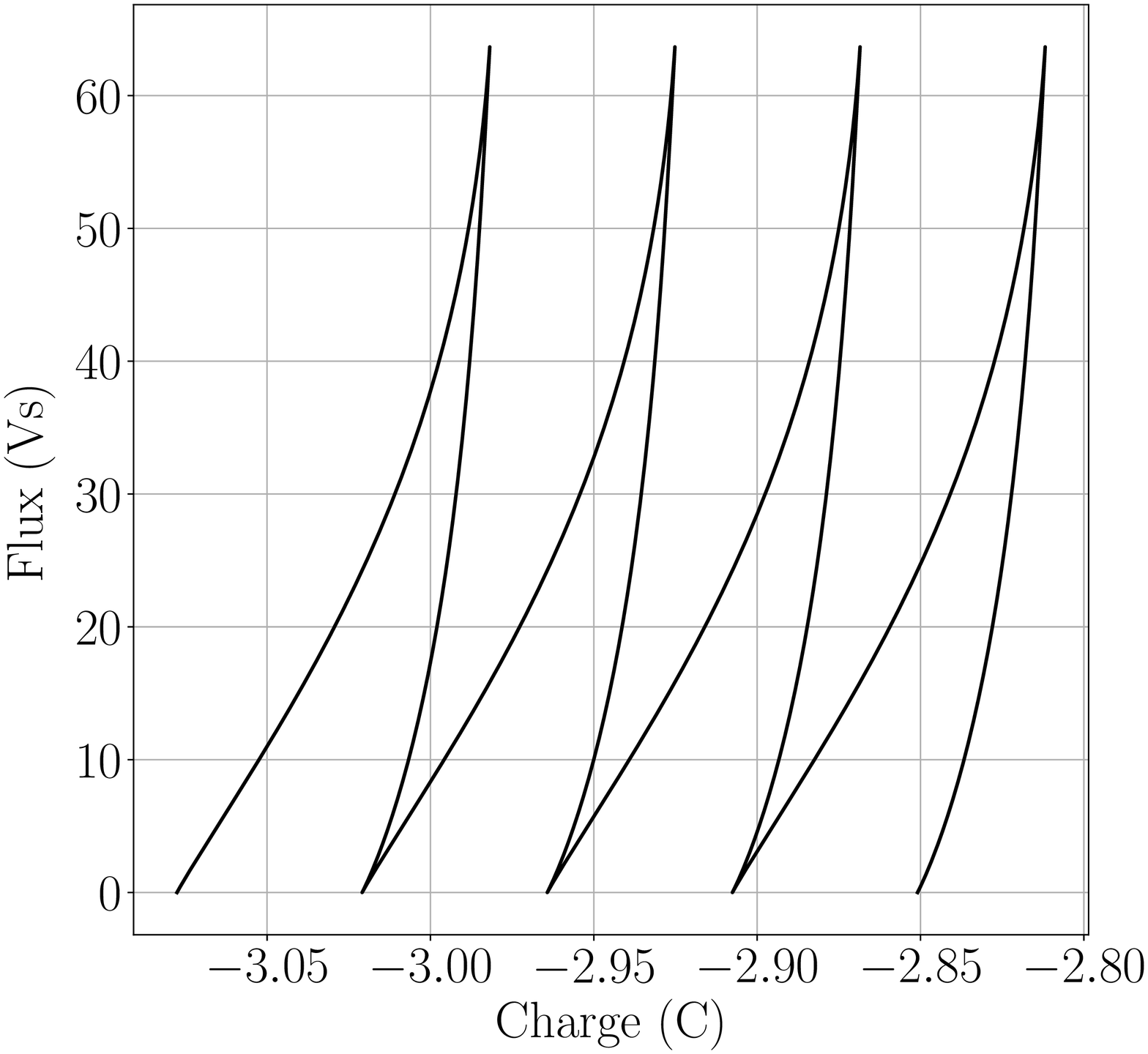} 
\caption{Axon model $\varphi-q$ curve for the following parameters:  
$f=50$~mHz, $g_K=36$~S, $E_K=10$~V, 
$\gamma=0.01$~V$^{-1}$s$^{-1}$, ${\hat V}=10$~V, $\tau=8$~s.} 
\label{pq_axon} 
\end{figure}

As a result of this lack of symmetry, we do not have a single-valued curve 
in the $\varphi-q$ plane, and in this case we do not have a complementary 
asymmetry as the one that exists in the case of the thermistor 
(Fig.~\ref{pq_medium}), therefore  
the curve in the $\varphi-q$ plane is not even closed: we show such a  
curve for 4 periods in Fig.~\ref{pq_axon}. As a consequence, at each cycle  
there is a net transfer of charge, with a value corresponding to the distance  
between two consecutive intersections with the horizontal axis. 
 
\begin{figure}
\centering
\includegraphics[scale=0.8]{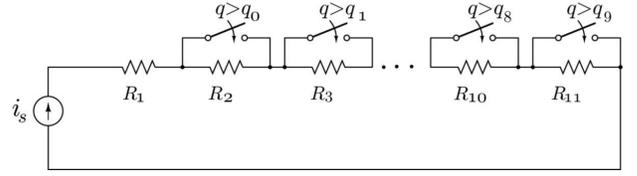}
\caption{Schematic representation of a memristor obtained by switching
resistors as a function of the
charge $q$.}
\label{circuito1}
\end{figure}

Let us now consider the circuit of Fig.~\ref{circuito1}, for which we   
assume charge dependent switches. As a consequence, we have an overall   
resistance depending on the charge $q$ that has flowed through the device.   
The current source $i_s$ provides the input signal, and $i_s$ is assumed to   
vary as $i_{s}(t)=A\sin(\omega t)$.   

The closing/opening sequences of the 10 switches are as follows:  
Closing sequence - switch 1 closes once the electrical charge $q$ that has  
flowed through $R_1$ satisfies $q\ge q_0=2A/11\omega$. 
The second switch closes   
when $q\ge 2q_0$, while the $i$th switch closes when $q\ge iq_0$, thus the   
the $10$th switch closes when $q\ge 10q_0$. At the end of the positive   
semi-period,   
a charge $q=11q_0$ has flowed through $R_1$, and all the switches   
are closed.   
 
Opening sequence - this sequence takes place in the negative semi-period,  
therefore the charge $q$ now decreases: the $10$th switch opens when   
$q\le 10q_0$, the $i$th switch opens when $q\le iq_0$, and switch 1 opens   
when $q\le q_0$. At the end of this sequence all switches are open and a new   
cycle can start.  
  
As a first case, we assume that all the resistors are time-independent.   
  
\begin{figure}
\centering
\includegraphics[scale=0.3]{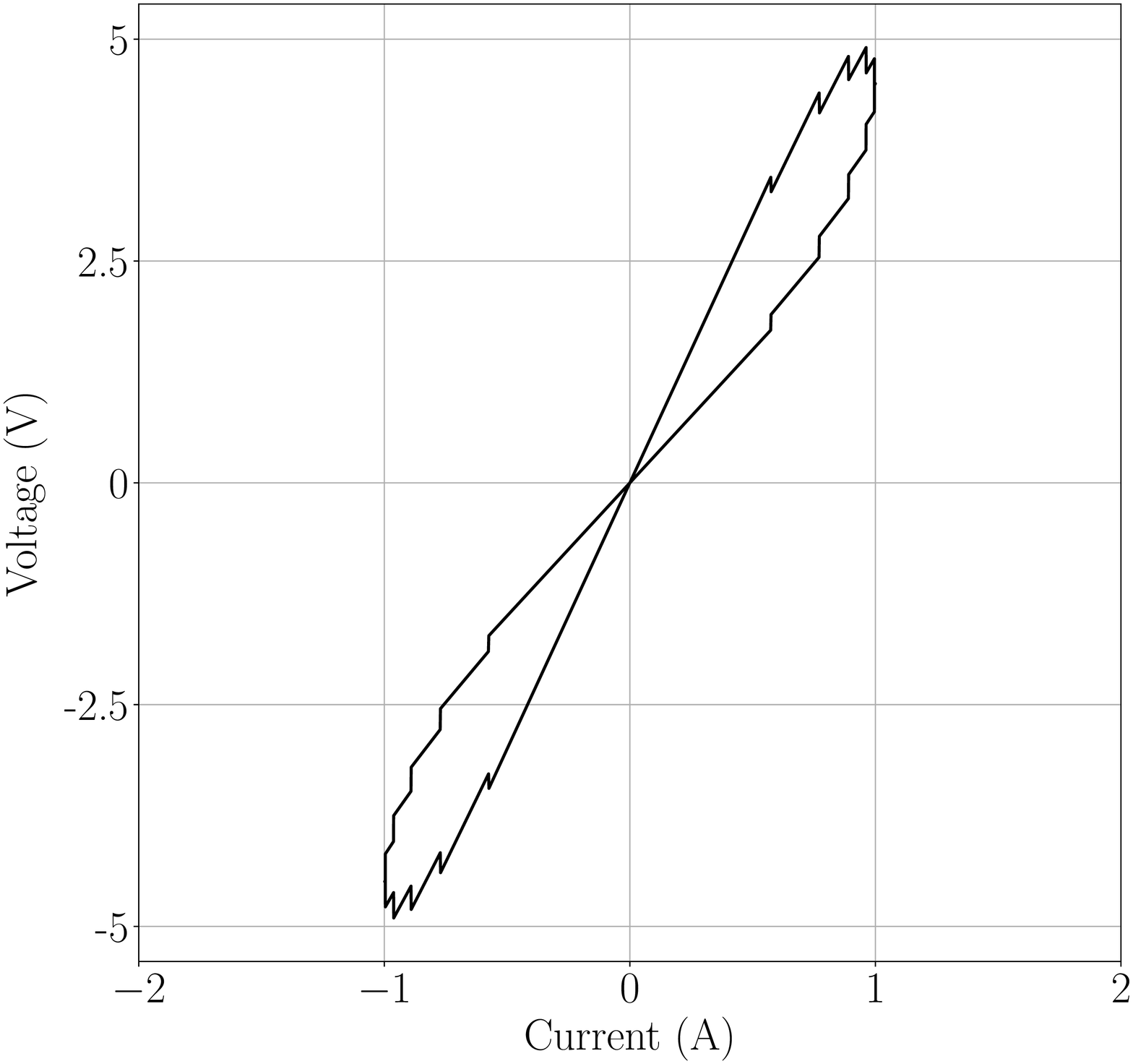}
\caption{Pinched hysteresis loop for the circuit of Fig.~\ref{circuito1}
with time-independent $R_1$, and $A=1$, $f=0.1$~Hz, $R_2=R_3=...=R_{11}=0.3~(\Omega)$,
$q_0=2A/(11\omega)$.}
\label{sim10}
\end{figure}

\begin{figure}
\centering
\includegraphics[scale=0.3]{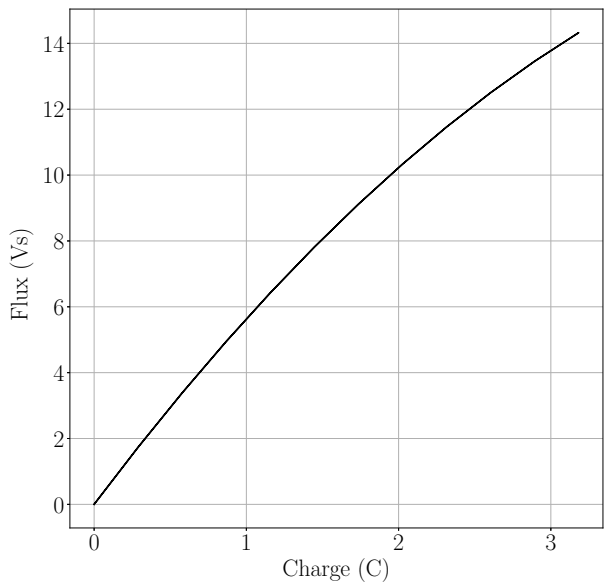}
\caption{$\varphi-q$ curve for the circuit of Fig.~\ref{circuito1}
with time-independent $R_1$, and $A=1$, $f=0.1$~Hz, $R_2=R_3=...=R_{11}=0.3~(\Omega)$,
$q_0=2A/(11\omega)$.}
\label{pqsim10}
\end{figure}

Simulation results are shown in Fig.~\ref{sim10}: we notice that the loop is   
pinched at the origin and is symmetric around the origin. Such a symmetry 
implies the single-valuedness of the $\varphi-q$ curve, which is reported 
in Fig.~\ref{pqsim10}.  
Therefore this simple model is compliant with the original requirement~\cite{chua_71}  
for a memristor, which includes the unambiguous definition through a  
curve in the $\varphi-q$ plane. 

\begin{figure}
\centering
\includegraphics[scale=0.25]{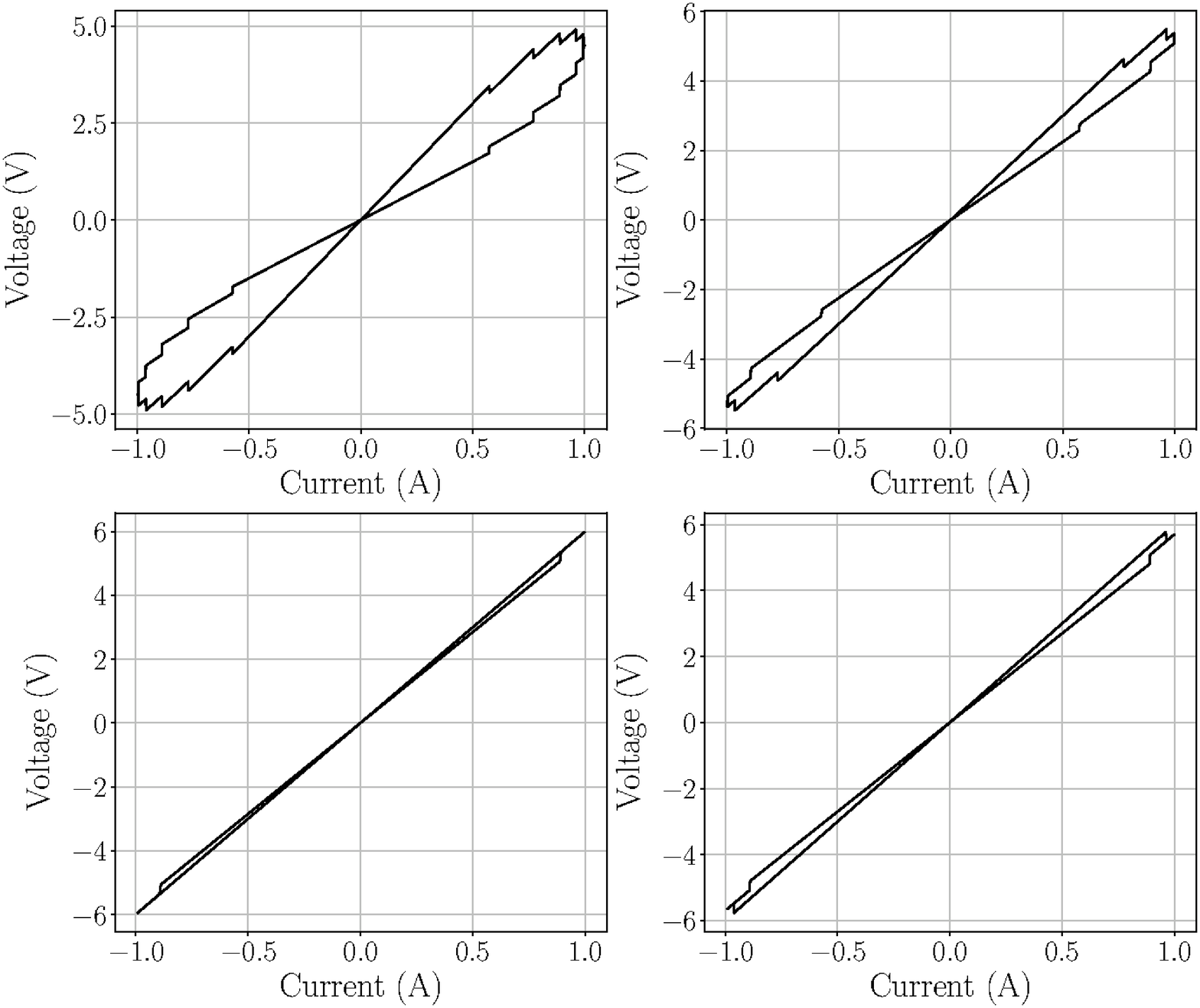}
\caption{Pinched hysteresis loops for the circuit of Fig.~\ref{circuito1}.
Proceeding clockwise from the top-left subplot the selected frequencies are:
$f=0.1$~Hz, $f=0.2$~Hz, $f=0.4$~Hz, $f=0.8$~Hz. The other parameters are the
same adopted for Fig.~\ref{sim10}.}
\label{final}
\end{figure}

The width of the   
loop in Fig.~\ref{sim10} depends on the frequency of the current.  
This is shown in Fig.~\ref{final}: proceeding clockwise from  
the top-left subplot we have: $f=0.1$~Hz, $f=0.2$~Hz, $f=0.4$~Hz, and  
$f=0.8$~Hz. We clearly observe how the loop area decreases for higher  
frequencies, for which the charge transfer is smaller, therefore the  
charge-dependent resistor variation decreases, too. With reference to  
Fig.~\ref{circuito1}, this means that a smaller number of switches is  
operated at higher frequencies.  

\begin{figure} 
\centering 
\includegraphics[scale=0.25]{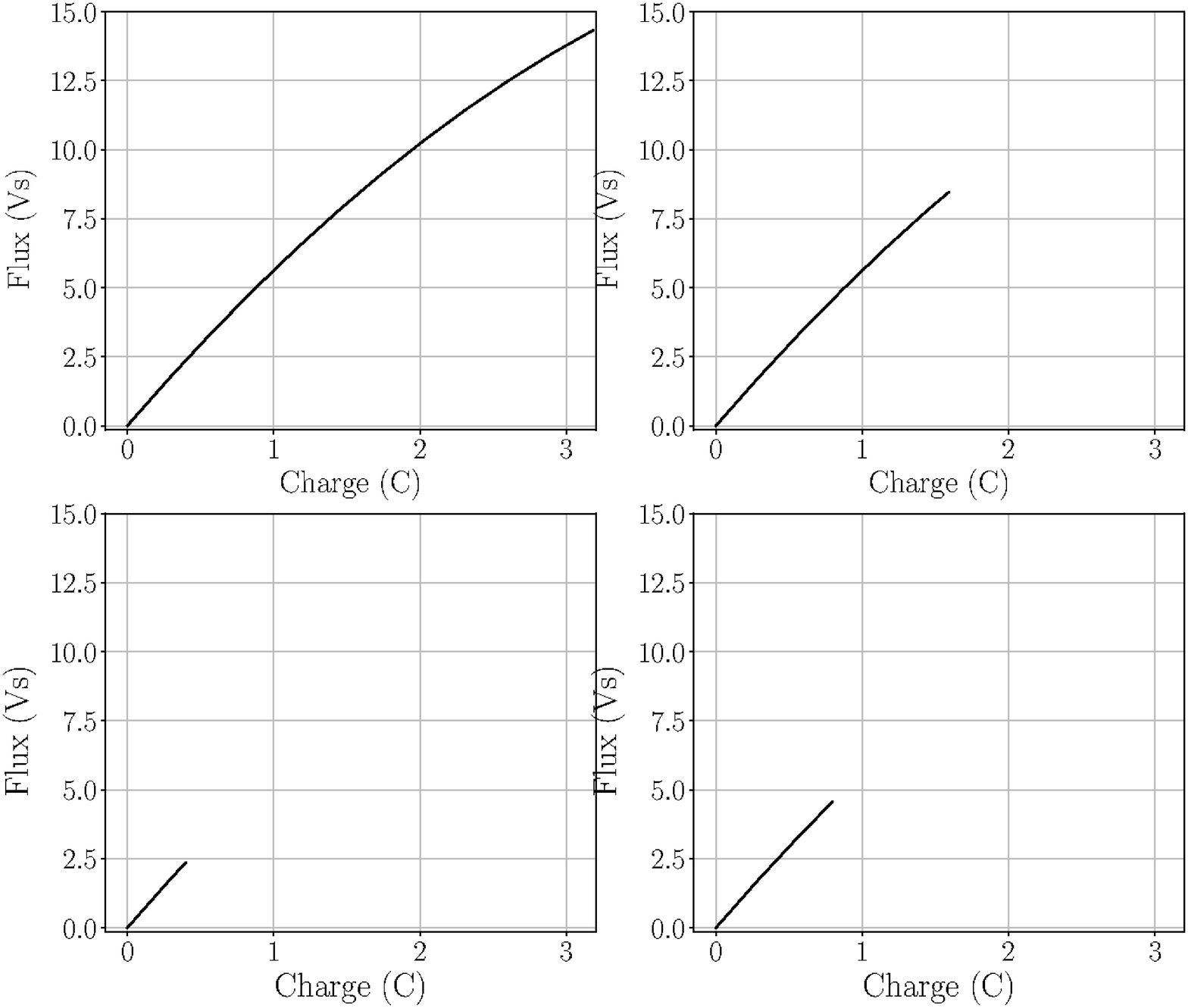}
\caption{$\varphi-q$ curves for the circuit of Fig.~\ref{circuito1}. 
Proceeding clockwise from the top-left subplot the selected frequencies are: 
$f=0.1$~Hz, $f=0.2$~Hz, $f=0.4$~Hz, $f=0.8$~Hz. The other parameters are the 
same adopted for Figs.~\ref{sim10},\ref{final}.} 
\label{pq_final} 
\end{figure}

The corresponding $\varphi-q$ curves are reported in Fig.~\ref{pq_final}, 
where they are arranged in a clockwise fashion as in Fig.~\ref{final}. 
  
\begin{figure}
\centering
\includegraphics[scale=0.30]{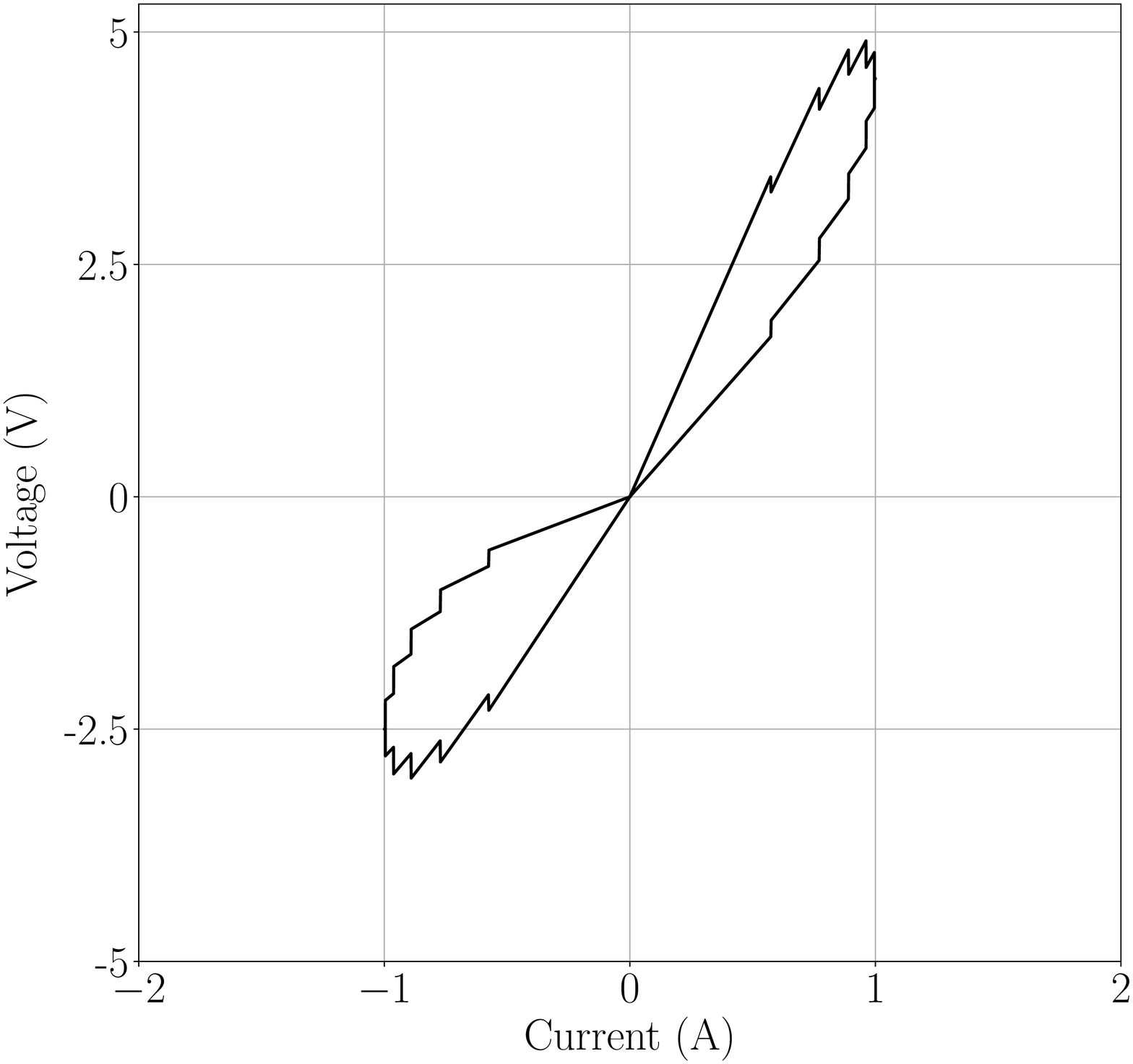}
\caption{Pinched hysteresis loop for the circuit of Fig.~\ref{circuito1} with
        time-dependent $R_1$, and $A=1$, $f=0.1$~Hz, 
        $R_2=R_3=...=R_{11}=0.3~(\Omega)$,
        $q_0=2A/(11\omega)$.}
\label{r1timevar}
\end{figure}

In the second case, $R_1$ is instead a resistor whose value has a periodic
dependence on time, according to:
\begin{equation}
\label{eq:aqui-le-mostramos-como-hacerle-la-llave-grande}
R_1(t) = \left\{
\begin{array}{ll}
3\,\,\Omega      & \mathrm{if\ } 0 < t \le T/2 \\
1\,\,\Omega & \mathrm{if\ } T/2 < t \le T \\
\end{array}
\right.
\end{equation}
Simulation results are shown in Fig.~\ref{r1timevar}: we can observe that
the $v-i$  curve is still pinched at the origin, as a result of the purely
resistive nature of the circuit. 

\begin{figure} 
\centering 
\includegraphics[scale=0.30]{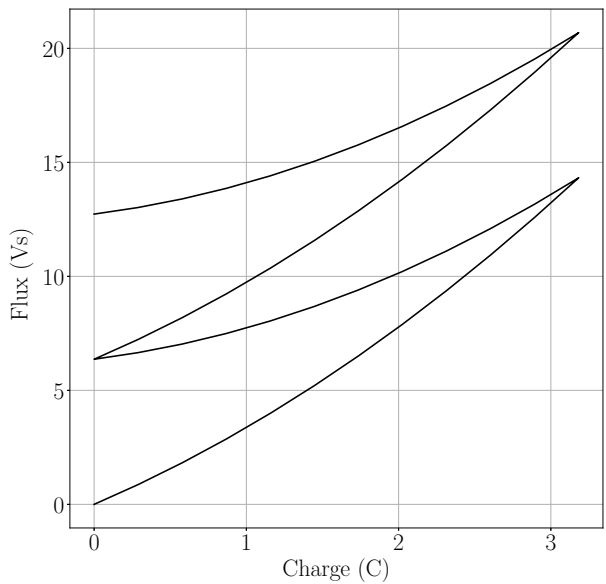} 
\caption{$\varphi-q$ curve for the circuit of Fig.~\ref{circuito1} with 
        time-dependent $R_1$, and $A=1$, $f=0.1$~Hz, 
        $R_2=R_3=...=R_{11}=0.3~(\Omega)$, 
        $q_0=2A/(11\omega)$.} 
\label{pqr1timevar} 
\end{figure}

In particular the average slope in the first quadrant is larger than that
in the third quadrant, as a consequence of the larger value of the resistance
during the positive semi-period.
In this case the $\varphi-q$ curve is neither single-valued nor closed,
because the hysteresis loop is completely asymmetric. We report it in
Fig.~\ref{pqr1timevar} for two consecutive cycles: after each cycle the
charge returns to zero since the driving quantity is a sinusoidal current,
while the flux linkage exhibits a net increase at each cycle since the
integral of the voltage in the positive semi-period of the current is
larger than that in the negative semi-period.

\begin{figure} 
\centering 
\includegraphics[scale=0.30]{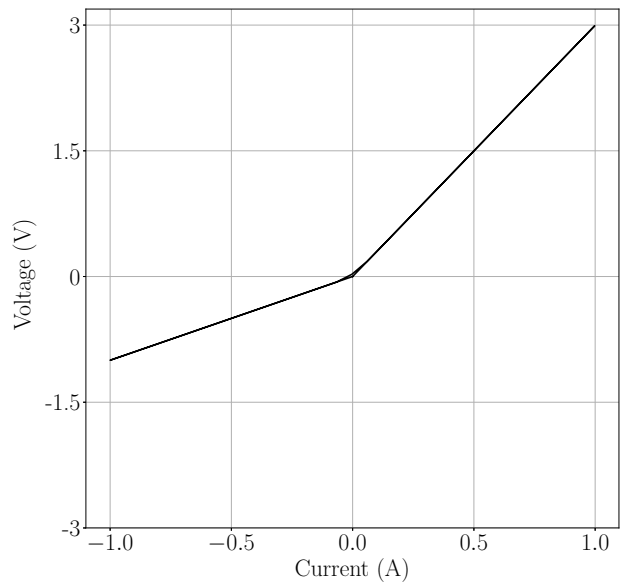} 
\caption{Pinched hysteresis loop for the circuit of Fig.~\ref{circuito1} with   
	time-dependent $R_1$, and $A=1$, $f=10$~Hz, $R_2=R_3=...=R_{11}=0.3~(\Omega)$,   
	$q_0=2A/(11\omega)$.} 
\label{r1timevarf10} 
\end{figure}

For this circuit, in the high-frequency limit, the $v-i$ hysteresis
loop collapses onto two segments, each with a slope equal
to the resistance of $R_1$ in the corresponding semi-period (see
Fig.~\ref{r1timevarf10}.)

We point out that both the memristor circuit with time independent
resistors and that with a time-dependent resistor
satisfy the passivity criterion adopted by Chua~\cite{chua_71},
i.e., that the instantaneous voltage to current ratio is always positive.
This remains
true also if the switches are controlled by circuits with active components
(such as transistors), because we can assume that the power needed to
operate them is obtained from the current flowing through the device.
%(with some minor change in the $v-i$ curves).
Indeed, this is what happens also in the memristor of \cite{strukov},
where vacancy migration is ``powered'' by the bias current.

\begin{figure}  
    \centering  
\includegraphics[scale=0.8]{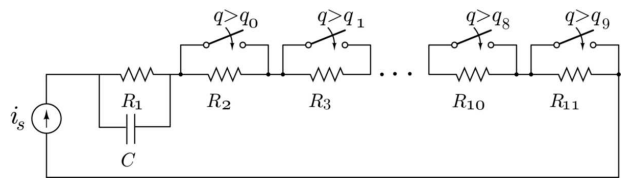}    
\caption{Circuit of Fig.~\ref{circuito1} with the inclusion of a   
capacitor element.}  
    \label{circuito3}  
\end{figure}

In the last simulation we included a capacitor connected in   
parallel to $R_1$, as shown in Fig.~\ref{circuito3}. 

\begin{figure}  
     \centering  
     \includegraphics[scale=0.3]{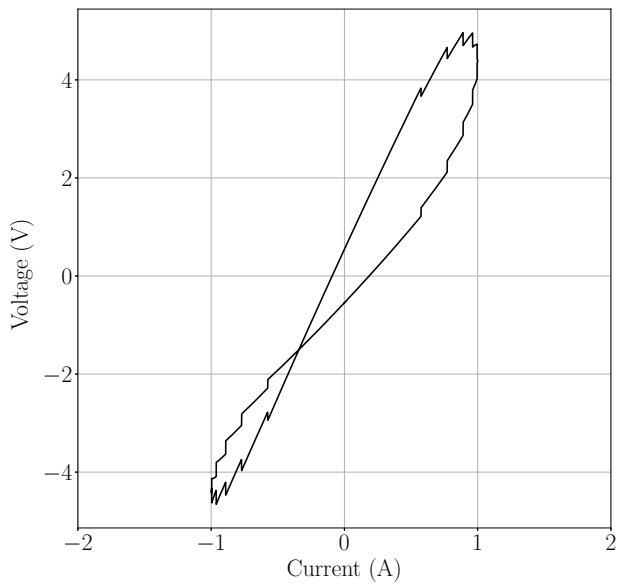}  
     \caption{Pinched hysteresis loop for a circuit consisting of a resistor   
depending on the charge $q$ and of a capacitor.}  
     \label{pinched3}  
\end{figure}  
Parameter values, in
particular $C=0.1~F$, are
chosen in such a way as to clearly show how the presence of a reactive
element implies that the $v-i$ curve is not pinched at the origin any longer,
and thus the resulting device is neither a memristor nor a memristive
system.
This is shown in Fig.~\ref{pinched3}.

\section{Conclusion}  
We have reconsidered the concept of memristor within   
circuit theory, starting from the definitions of basic circuit  
elements, in order to understand its actual nature and whether it is truly a 
fourth basic circuit element. 
 
As a major aspect of our analysis, we have reached the conclusion that 
the memristor is not a fourth basic circuit element on the same  
footing as the traditional basic elements, while it can rather 
be interpreted as an extension of the resistor concept.  
 
In particular, we have shown that, as long as the input and output 
quantities are a voltage and a current, a memristive system can be 
seen as a generalized time-variant resistor (generalized in the sense 
that, with respect to the definition provided by Mauro in 1961, also  
an additional direct dependence on time is allowed). 

We have also stressed the relevant point that the meaningful quantity 
in the memristor definition is the flux linkage 
$\varphi$, to be intended as the time integral
of the voltage and not as the actual magnetic flux. Without this clarification, 
ambiguities may appear and conclusions that are not fully sound may be drawn.
 
Motivated by the presence in the recent memristor literature of references 
to systems studied in the 50's and 60's (such as the thermistor and 
the giant squid axon), in which an  
anomalous reactive behavior had been observed, we have revisited the  
original papers, realizing that such anomalous behavior had been understood 
and explained at the time with concepts that we can use to provide a general 
explanation of the nature of the memristor. 
At the same time we have analyzed the evolution of the definition of the  
memristor and of the memristive systems, starting from the original 1971 
formulation and its 1976 extension.

In particular, we have analyzed the operation of a recently proposed
``$\Phi$-memristor,'' pointing out that its $v-i$ curve cannot be 
rigorously pinched in the origin, because coupling with the magnetic
flux involves the unavoidable presence of an inductance.
 
Finally, we have presented a few numerical examples of memristive systems and 
memristors, showing that, in order to have a single-valued curve in the 
$\varphi-q$ plane, the $v-i$ curve must possess particular symmetry properties,
which are present only in the case of the basic time-invariant memristor as 
defined by Chua in 1971. Such an essential property 
is not characteristic of generic memristive systems.

We believe that we have contributed to a clarification of the memristor 
concept that can be useful for the memristor community, in order to further
develop applications and to better focus on new and original problems.
  
\section*{Acknowledgment}  
M.B. acknowledges financial support from UTA Mayor project No. 8765-17.  
M.M. acknowledges partial support by the Italian Ministry of Education   
and Research (MIUR) in the framework of the CrossLab project (Departments   
of Excellence).  
  
\ifCLASSOPTIONcaptionsoff  
  \newpage  
\fi

\end{document}